\documentclass[fleqn,usenatbib]{mnras}

\usepackage{newtxtext,newtxmath}

\usepackage[T1]{fontenc}
\usepackage{pdflscape}
\DeclareRobustCommand{\VAN}[3]{#2}
\let\VANthebibliography\thebibliography
\def\thebibliography{\DeclareRobustCommand{\VAN}[3]{##3}\VANthebibliography}

\usepackage{graphicx}	
\usepackage{amsmath}	

\title[Stellar population at $z\sim$8]{Resolving Ambiguities in the Inferred Star Formation Histories of Intense [O III] Emitters in the Reionisation Era }

\author[N. Laporte et al.]{
N. Laporte,$^{1,2}$\thanks{E-mail: nl408\@ cam.ac.uk}
R. S. Ellis,$^{3}$
C. E. C. Witten,$^{4}$
G. Roberts-Borsani, $^{5}$
\\
$^{1}$Kavli Institute for Cosmology,University of Cambridge, Madingley Road, Cambridge CB3 0HA, UK\\
$^{2}$Cavendish Laboratory,University of Cambridge,19 JJ Thomson Avenue, Cambridge CB3 0HE, UK\\
$^{3}$ Department of Physics and Astronomy, University College London, Gower Street, London WC1E 6BT, UK \\
$^{4}$Institute of Astronomy,University of Cambridge, Madingley Road, Cambridge CB3 0HA, UK\\
$^{5}$Department of Physics and Astronomy, University of California, Los Angeles, CA 90095, USA \\
}

\date{Accepted XXX. Received YYY; in original form ZZZ}

\pubyear{2022}

\begin{document}
\label{firstpage}
\pagerange{\pageref{firstpage}--\pageref{lastpage}}
\maketitle

\begin{abstract}
Early JWST spectroscopic campaigns have confirmed the presence of strong [O III] line-emitting galaxies in the redshift interval $7<z<9$. Although deduced earlier from \textit{Spitzer} photometry as indicative of young stellar populations, some studies suggested the relevant photometric excesses attributed to [O III] emission could, in part, be due to Balmer breaks arising from older stars. We demonstrate that this is likely the case by exploiting medium-band near-infrared JWST photometry in the \textit{Hubble} Ultra Deep Field. We locate a sample of 6 galaxies with redshifts 8.2$<z<$8.6 for which the relevant medium-band filters enables us to separate the contributions of [O III] emission and a Balmer break, thereby breaking earlier degeneracies of interpretation. The technique is particularly valuable since it provides photometric redshifts whose precision, $\Delta\,z\simeq\,\pm0.08$, approaches that of spectroscopic campaigns now underway with JWST. Although some sources are young, a third of our sample have prominent Balmer breaks consistent with stellar ages of $\geq$150 Myr. Our results indicate that even intense [O III]
emitters experienced episodes of earlier star formation to $z\sim$10 and beyond, as is now being independently deduced from direct detection of the progenitors of similar systems.

\end{abstract}




\section{Introduction}
The spectroscopic and photometric capabilities of the recently-commissioned \textit{James Webb} Space Telescope (JWST) are transforming our ability to study the detailed physical properties of galaxies in the reionisation era. Extending the photometric coverage of \textit{Hubble} Space Telescope (HST) beyond 1.6 microns enables measures of the rest-frame optical and near-infrared fluxes of high redshift galaxies (\citealt{2022ApJ...938L..15C}, \citealt{2022arXiv220709434N}, \citealt{2022arXiv220712356D}, \citealt{2022arXiv220801612H}),  whereas the wide wavelength spectroscopic coverage can reveal gas-phase metallicities and measures of the ionizing radiation field necessary to pinpoint early AGN activity (\citealt{2022A&A...665L...4S}, \citealt{2022arXiv220712388T}, \citealt{2022MNRAS.tmp.2550C}, \citealt{2022MNRAS.tmp.2470K}).

Although much attention has focused on the challenges of using JWST to penetrate beyond the redshift $z\sim$10 horizon of HST (e.g. \citealt{2022arXiv220802794N}), in this paper we attempt to resolve a related ambiguity that concerns the star-formation histories of galaxies at lower redshifts $z\sim$7-9. If, as proposed from early JWST discoveries, there is a detectable population of star forming galaxies extending over $z\sim$10-16, there should be well-established stellar populations with ages of $\sim$200-400 Myr in these later galaxies.

The challenge of deducing stellar ages of star-forming galaxies in the reionisation era, as opposed to simply conducting a redshift-dependent census (e.g. \citealt{2016MNRAS.459.3812M}, \citealt{2016ApJ...819..129O}) has been well-documented in the pre-JWST literature. A fundamental issue is locating old stars through spectral energy distribution (SED) fitting in systems where recent star formation obscures their presence (\citealt{2022arXiv220801599W}, \citealt{2022MNRAS.513.5211T}). The most promising index of the age of a stellar population is the Balmer break whose utility originates via the dominance of atomic hydrogen as the source of opacity in the atmospheres of A-type stars whose main sequence lifetimes are $\sim$1 Gyr \citep{2015ApJ...800..108S}. Excess signals seen in the 3.6 and 4.5 micron IRAC bands for $z\sim$6-8 galaxies observed with the \textit{Spitzer} Space Telescope have been conventionally interpreted as arising from intense [O III] emission associated with younger stellar systems (\citealt{2013ApJ...777L..19L}, \citealt{2014ApJ...784...58S}). However, others have argued a significant contribution to these excess fluxes could arise from a Balmer break \citep{2020MNRAS.497.3440R}. Only beyond a redshift of $z\sim$9 does this ambiguity disappear when [O III] emission is redshifted beyond the 4.5 micron band. \textit{Spitzer} studies of a modest sample of $z>$9 galaxies (\citealt{2018Natur.557..392H}, \citealt{2021MNRAS.505.3336L}) reveal the presence of Balmer break supporting earlier star formation up to redshift $z\sim$15 as now claimed by early JWST observations \citep{2022arXiv220712356D}.

In this paper we return to resolving the ambiguity in the interpretation of the spectral energy distributions of galaxies at $z\sim$7-9 for which rich photometric datasets are now emerging from JWST imaging. The use of medium-band photometric filters in the NIRCam instrument offers the potential to isolate the contribution of [O III] emission from that of the Balmer break as well as to provide photometric redshifts of good precision \citep{2021ApJ...910...86R}. We apply this technique to a recently-released photometric dataset in the \textit{Hubble} Ultra Deep Field (UDF). Although a modest sample at this early stage of JWST observations, we demonstrate the promising ability to separate genuine young $z\sim 8.2-8.6$ systems from those with significant earlier star formation histories. Although such age discrimination is also possible through diagnostic spectroscopy, we argue that more extensive photometric studies of this nature will lead to a complementary understanding of the history of star formation in the first 400 Myr of cosmic history.

A plan of the paper follows. In \S2 we discuss the UDF observational dataset, its processing and sample selection. \S3 presents the SED fitting that enables us to quantitatively separate the contributions of [O III] emission and the age-dependent Balmer break. We discuss the implications of our results in terms of early star formation in \S4 and our conclusions are presented in \S5.

\section{Observational Data and Sample Selection}
The \textit{James Webb} Space Telescope observed the UDF field (RA=03:32:38, DEC=-27:47:00) on October 11 and 12 2022 with NIRCam in 5 medium-band filters (F182M, F210M, F430M, F460M and F480M) as part of a Cycle 1 GO program (\# 1963 - P.I. : Christina Williams). The exposure time in each filter was $\sim$4hrs for F430M and F460M, and $\sim$ 8hrs for F182M, F210M and F480M filters. We downloaded the uncalibrated data from the MAST archive and processed these using version 1.7.2 of the \textit{JWST} pipeline 
incorporating the latest available calibration files (jwst\_1009.pmap - updated on October 26$^{th}$ 2022). We follow a 3-step procedure: the Stage 1 detector processing (also known as "ramps-to-slopes” processing - calwebb\_detector1), Stage 2 which calibrates the slope images and Stage 3 which creates the final mosaic. We also took extra care to measure and remove the horizontal and vertical striping from the two count rate images produced after Stage 1. Photometric catalogues were constructed using the \texttt{Source Catalog} step of the pipeline with the following parameters : \texttt{snr\_threshold}=2.0$\sigma$, \texttt{npixels}=10.0 pixels, \texttt{kernel\_fwhm}=1.990, 2.304, 2.300, 2.459 and 2.574 respectively for F182M, F210M, F430M, F460M and F480M,  and \texttt{deblend}=True.  

\begin{figure}
    \centering
    \includegraphics[width=0.45\textwidth]{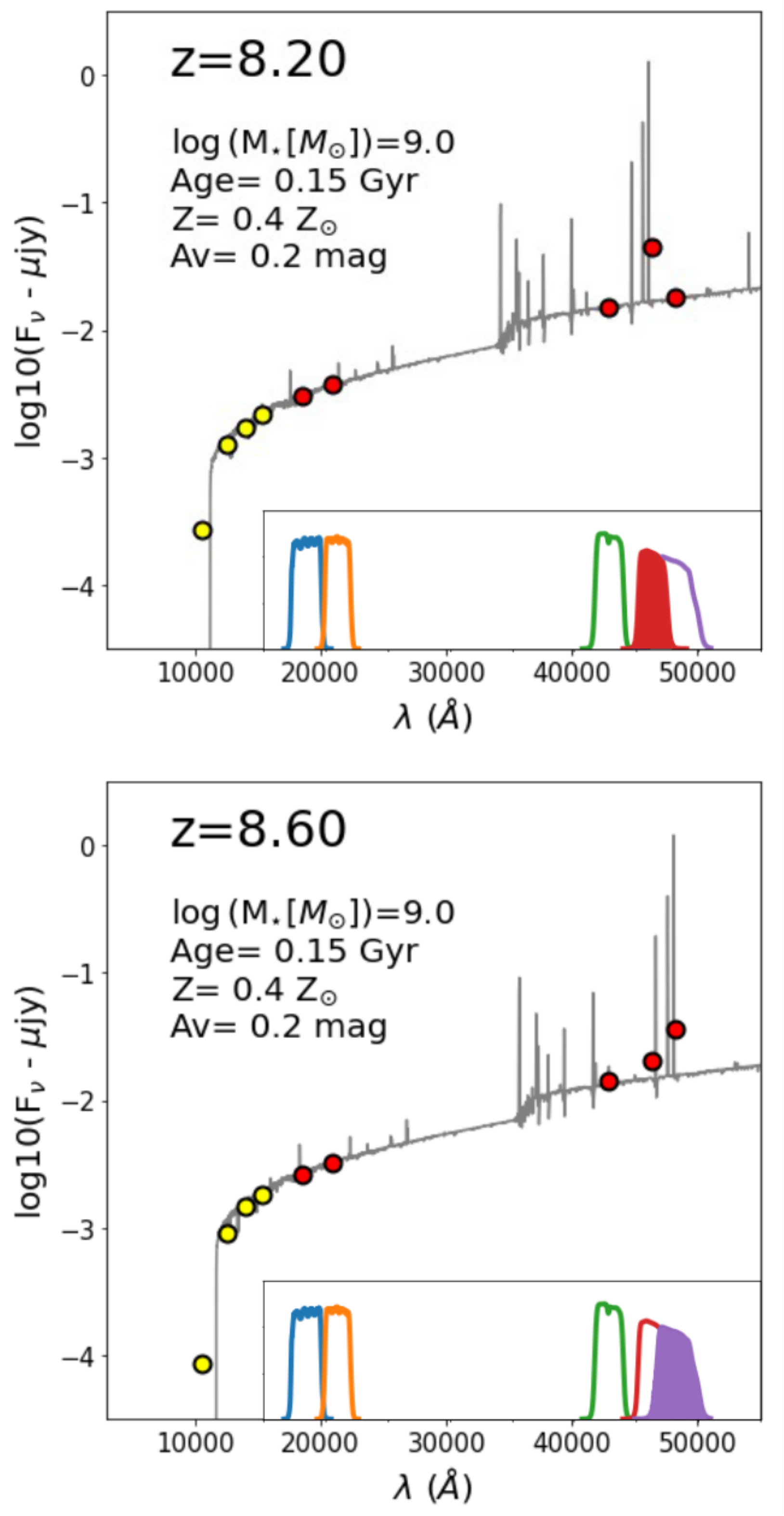}
    \caption{Examples of a simulated SED for a galaxy at $z=8.20$ (top) and  $z=8.60$ (bottom). Yellow dots show expected fluxes from HST, red dots are the expected fluxes in the NIRCam images.  The parameters of the input models (stellar mass, dust reddening, metallicity and age) are listed. Insert panels show the transmission curves of the medium-band filters, the filled curve indicates the filter used to select strong [OIII]5007 emitters at $z\geq$8. }
    \label{fig:bandwidth}
\end{figure}

\begin{figure*}
    \centering
    \includegraphics[width=1.0\textwidth]{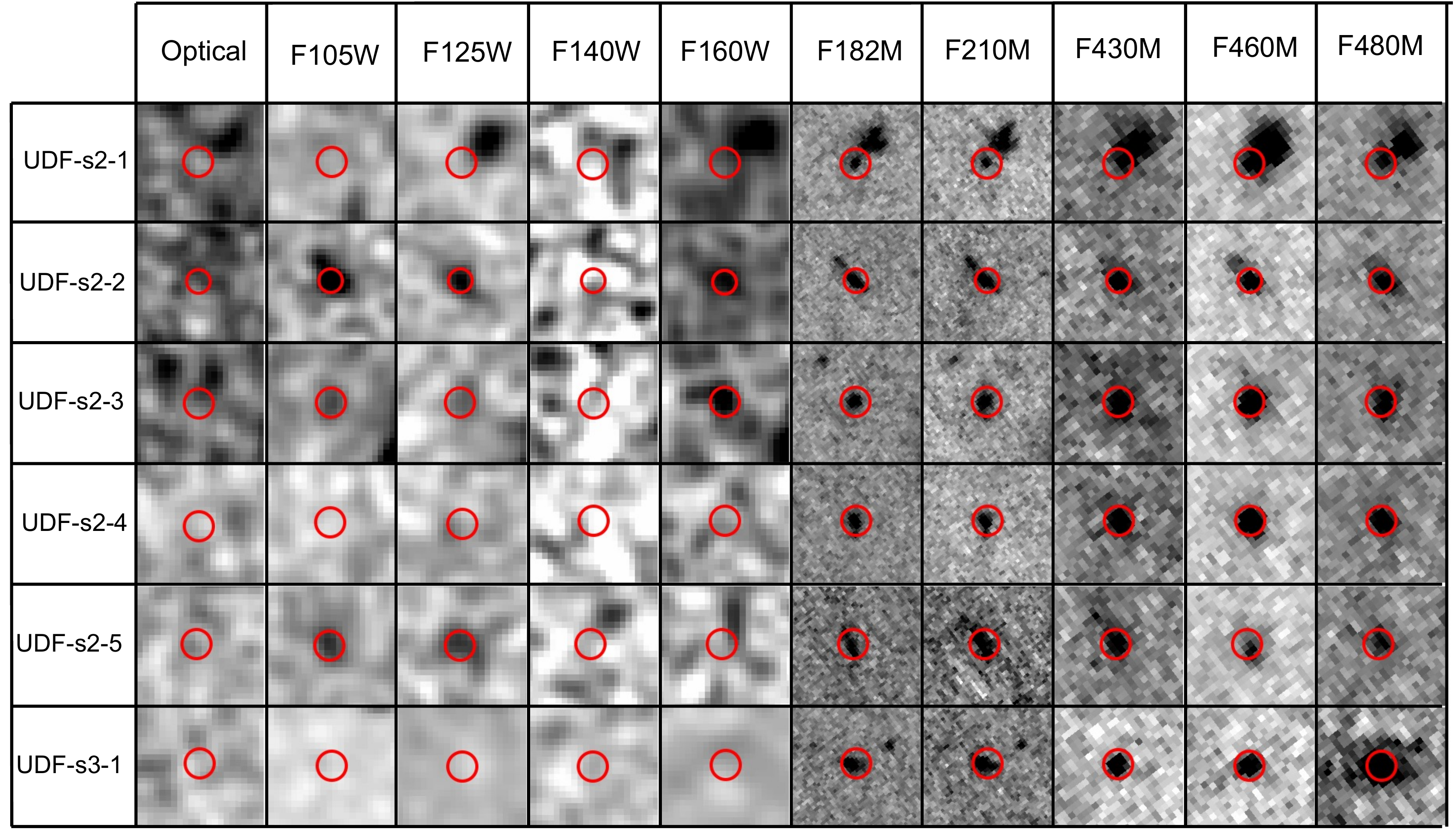}
    \caption{Image stamps of candidates for the $z\sim$8.2 (UDF-s2-*) and $z\sim$8.6 samples (UDF-s3-1). The position of each candidate is displayed by a red 0.3\arcsec diameter circle. The optical image represent the stack of the ACS F435W, F606W, F775W, F814W and F850LP images which have been smoothed for clarity.}
    \label{fig:stamps}
\end{figure*}

\begin{figure*}
    \centering
    \includegraphics[width=0.33\textwidth]{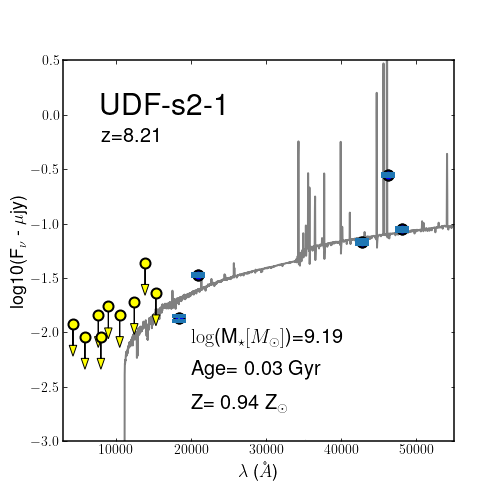}
    \includegraphics[width=0.33\textwidth]{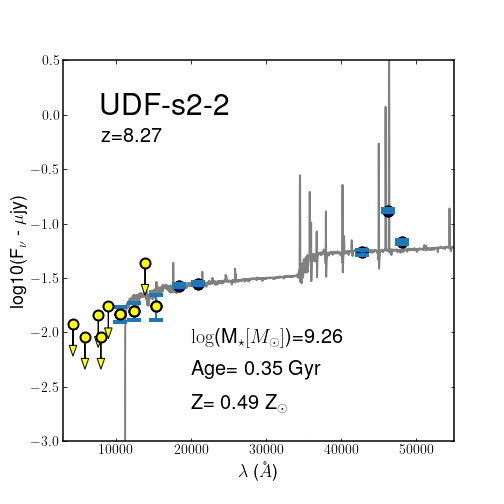}
    \includegraphics[width=0.33\textwidth]{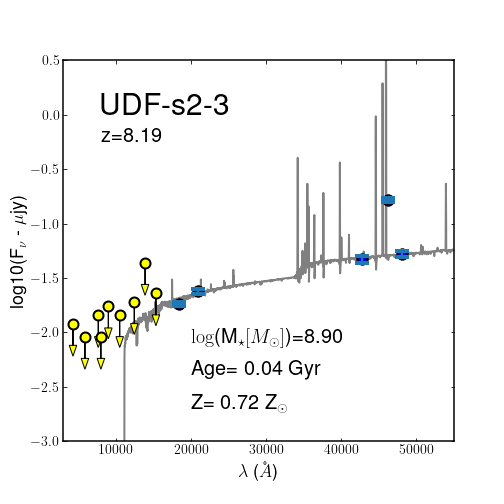} \\
    \includegraphics[width=0.33\textwidth]{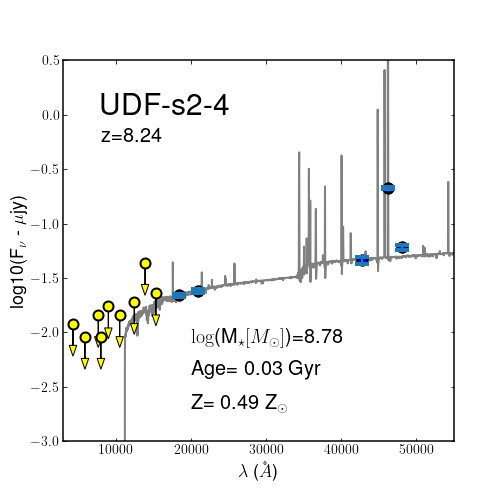} 
    \includegraphics[width=0.33\textwidth]{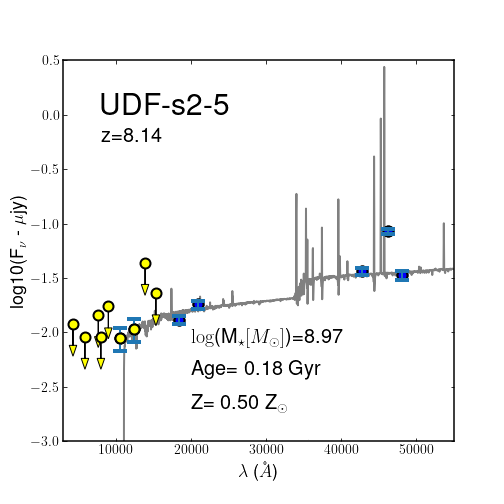}
    \includegraphics[width=0.33\textwidth]{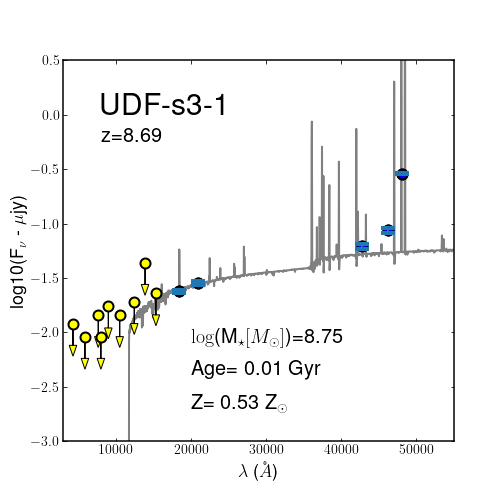} 

    \caption{Spectral Energy Distribution of the [OIII]5007\AA\ line emitters  in this study. Blue dots show  photometry from the NIRCam medium-band filters, yellow dots are from HST/ACS and HST/WFC3. Upper limits are at 2$\sigma$. The properties indicated in each panel are the parameters of the model plotted in black.}
    \label{fig:SED}
\end{figure*}

Our main goal is to constrain the age of the stellar population for $z\geq$8 galaxies whose photometric redshift can be estimated to good precision using the medium-band filters. At $z\geq$8, the brightest optical emission line system is [OIII]4959,5007 \AA\ (eg., \citealt{2013ApJ...777L..19L}, \citealt{2014ApJ...784...58S}) redshifted to $\lambda \geq$ 4$\mu$m. The UDF survey offers 3 medium-band filters at $\lambda \geq$ 4$\mu$m, namely F430M, F460M and F480M, the latter two allowing us to identify [OIII] at $z\sim$8.2 and 8.6 (Figure ~\ref{fig:bandwidth}). Therefore, our first selection criterion requires a $>$0.5 mag excess in either F460M or F480M to locate the [OIII] emission. In order to obtain estimates of the mass and age of the stellar population, we also require the object to be clearly detected ($\geq$5$\sigma$) in the other two $\lambda \geq$4$\mu$m filters unaffected by line emission.  Since we aim to estimate the fraction of [OIII]-emitting galaxies with a mature stellar population, we do not impose any constraint on the size of the Balmer break inferred from the (F210M - F430M) colour.  However, to avoid contamination by low-$z$ galaxies exhibiting strong H-$\alpha$, we combine our NIRCam data with ACS/HST photometry from the 3DHST survey (\citealt{2014ApJS..214...24S}, \citealt{2012ApJS..200...13B}). To ensure no foreground contamination, we impose a non-detection in the stacked optical ACS images at the 2 sigma level which serves to exclude $z\leq$6.5 galaxies. We also exclude dusty low-$z$ galaxies by removing sources with a red UV-slope, i.e. $m_{F182M}-m_{F210M} <$0.2. After visual inspection, the $z\sim$8.2 sample comprises 5 galaxies and the $z\sim$8.6 has one galaxy. Our sample is listed in Table~\ref{tab:properties} with the detailed photometry in Table~\ref{tab:photometry}. Image stamps and SEDs of our candidates are shown in Figure~\ref{fig:stamps} and Figure~\ref{fig:SED}, respectively.

\begin{figure}
    \centering
    \includegraphics[width=0.5\textwidth]{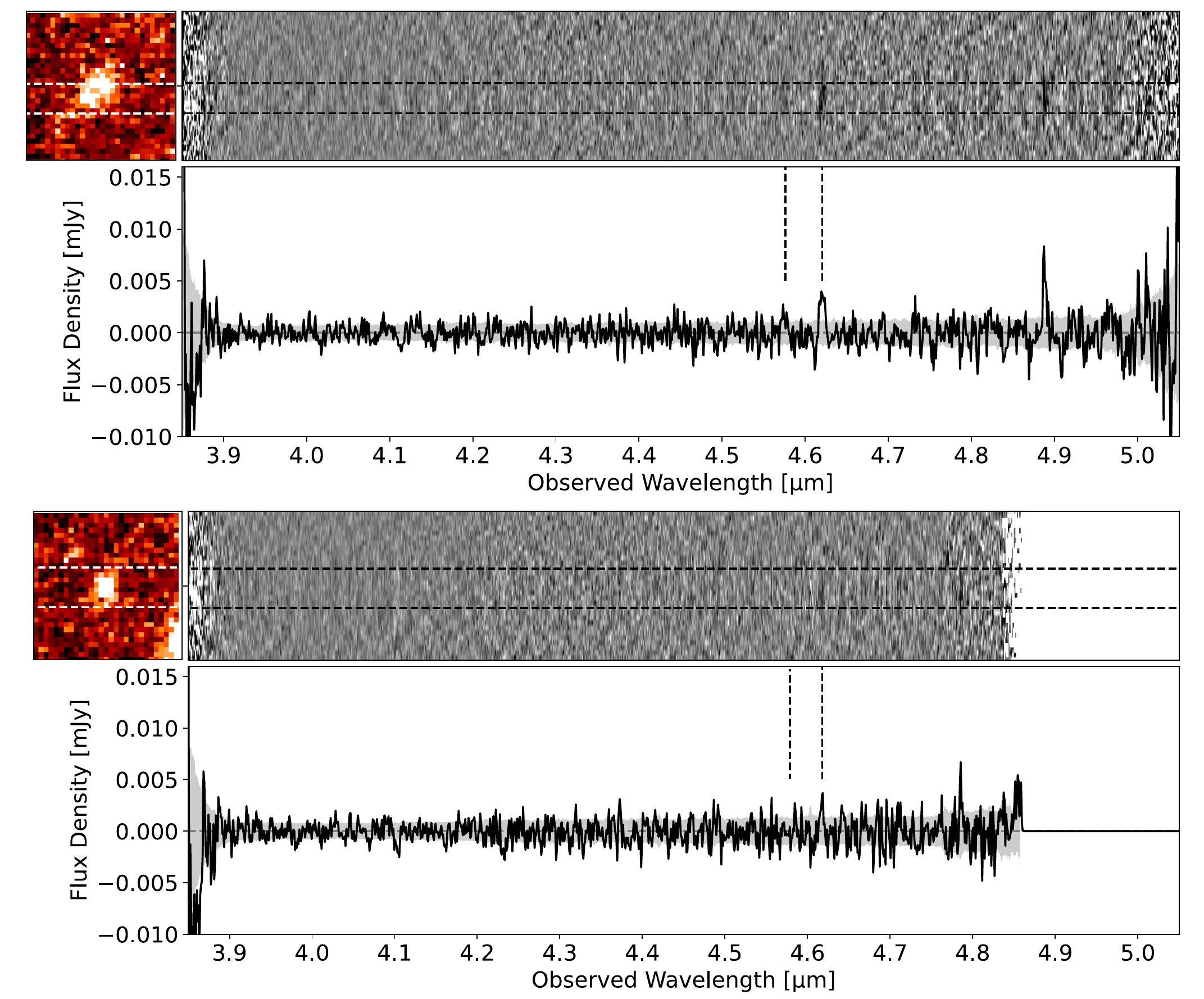}
    \caption{2D and 1D spectra of UDF-s2-1 (top) and UDF-s2-3 obtained with NIRCam/grism as part of the FRESCO survey (ID: 1895, PI: P. Oesch). A strong emission line is seen in both spectra at 4.6205$\mu$m (UDF-s2-1) and 4.6185$\mu$m (UDF-s2-3) consistent with [OIII]5007 at $z=8.228$ and $z=8.223$ respectively. This hypothesis is also supported by the detection of a weaker line at the expected position of [OIII]4959 for UDF-s2-1. }
    \label{fig:spectro}
\end{figure}

Spectroscopic follow-up of our sample has been undertaken by the FRESCO JWST survey (ID: 1895, PI: P. Oesch) which studied part of the UDF with the NIRCam grism at $\lambda \sim$4$\mu$m. The total integration time is 35.5 hours leading to a 6-sigma line sensitivity of $3.3 \times 10^{-18}$ erg/s/cm$^2$. The FRESCO survey utilizes only one of the two grisms of each NIRCam module, therefore the observed spectra of objects are subject to contamination from sources along the direction of dispersion of said grism (grism R). The data was reduced following the methods described in \citet{Sun+22} including flat fielding, background subtraction, 1/f noise subtraction and WCS assignment
We also perform a careful astrometric analysis to correct for any offsets between the short wavelength channel imaging (F182M and F210M) and long wavelength channel Grism spectra (F444W). We additionally perform a continuum subtraction from the 2D spectra of our sources in order to remove the effects of contaminating sources. Following the reduction we find that three objects from our sample were observed over the appropriate wavelength range (UDF-s2-1, UDF-s2-2 and UDF-s2-3) and three were observed across a restricted wavelength range that does not include the [OIII]4959,5007 emission lines (UDF-s2-4, UDF-s2-5 and UDF-s3-1). For UDF-s2-2, no emission line is detected suggesting an upper limit on the EW([OIII]5007\AA\ ) of 1870\AA\ at 3$\sigma$, consistent with our conclusion that this object has the smallest EW([OIII]4959,5007+H$\beta$) of our sample. UDF-s2-1, the brightest object in our sample, shows two strong emission lines at $\lambda$=4.8865$\mu$m and $\lambda$=4.6205$\mu$m and one weaker emission line that is visible in the extracted 1D spectrum at $\lambda$=4.5775$\mu$m. The two strong lines are slightly offset along the Y-axis suggesting that they arise from spatially-distinct objects whose spectra overlap (Figure\ref{fig:spectro}). Assuming that the line at $\lambda$=4.6205$\mu$m is [OIII]5007, the spectroscopic redshift would be $z=$8.228, as expected from our data. The weaker emission line identified at $\lambda$=4.5775$\mu$m would therefore be associated with the [OIII]4959 line. Although the spectrum of UDF-s2-3 is similarly contaminated, there is a clear emission line at $\lambda$=4.6185$\mu$m which, if [OIII]5007, would yield a spectroscopic redshift of $z=$8.224, also in agreement with our photometric value (see Table~\ref{tab:properties}).

\section{Physical properties}

The physical properties of our selected galaxies have been estimated following the method described in \citet{2021MNRAS.505.3336L} using \texttt{BAGPIPES} \citep{2018MNRAS.480.4379C} assuming several possible star formation histories (SFH - burst, constant, delayed and a combination of a recent burst with a constant star formation). The physical properties are then deduced from the fit with a maximum likelihood that incorporates the appropriate number of parameters for the SFH. Not surprisingly, \textit{BAGPIPES} identifies that strong [OIII]5007\AA\ emission lines are present in one of the medium-band filters. This leads to impressive precision for the photometric redshifts of our sample with an average uncertainty of $\Delta z_{phot}$=0.08. As a sanity check, we also force a $z\sim$6 solution for all our objects. At this redshift, the strong line seen in either F460M or F480M would be H$\alpha$ instead of [OIII]5007\AA . Comparing the $\log(ev)$ of the forced $z\sim$6 solutions with the best-fit $z\geq$8 solution clearly shows that the highest redshift solution is preferred for all our candidates, confirming that the line detected at $\lambda \geq$4$\mu$m is [OIII]+H$\beta$ at $z\geq$8. Moreover the $z\sim$6 solutions require a significant amount of dust ($A_v\geq$2mag).
Since none of our selected galaxies has been detected at 1.2mm in
the ALMA HUDF spectroscopic survey \citep{2016ApJ...833...72B}, we
consider such lower redshift explanations to be very unlikely. One candidate (UDF-s2-4) show similar likelihood for both fits, with a slightly preferred solution at $z\geq$8. 

\begin{figure}
    \centering
    \includegraphics[width=0.5\textwidth]{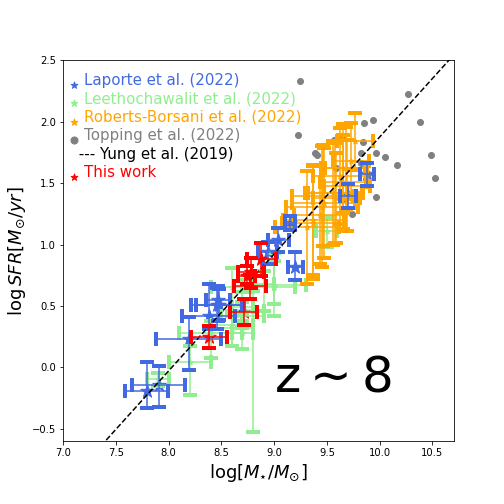}
    \caption{Star formation rate as a function of the stellar mass for our $z\sim$8 galaxies. Use of the medium-band filters has led to improved stellar masses (red dots) compared to other studies. We also plot constraints at $z\geq$8 from the GLASS \citep{2022arXiv220711135L} and BoRG \citep{2022ApJ...927..236R} surveys, the ALMA Rebels survey \citep{2022MNRAS.516..975T} and from a lensing field \citep{2022A&A...667L...3L}. The trend matches predictions at $z\sim$8 from simulations (black dotted line) \citep{2019MNRAS.490.2855Y} }
    \label{fig:main-sequence}
\end{figure}

The medium-band JWST data at $\lambda \geq$3$\mu$m also permits a fairly precise measure of the stellar mass, with an averaged uncertainty of $<\Delta \log [M_{\star}]>$=0.21. Figure \ref{fig:main-sequence} show the distribution of our 6 sources in the M$_{\star}$ vs SFR diagram. The stellar masses probed in this study range from 10$^8$ to 10$^9$ M$_{\odot}$, consistent with that deduced in other early JWST datasets (e.g. \citealt{2022arXiv220711135L}, \citealt{2022A&A...667L...3L}). The tight correlation seen with the star formation rate agrees well with recent simulations by \citet{2019MNRAS.490.2855Y}. 

\begin{figure}
    \centering
    \includegraphics[width=0.5\textwidth]{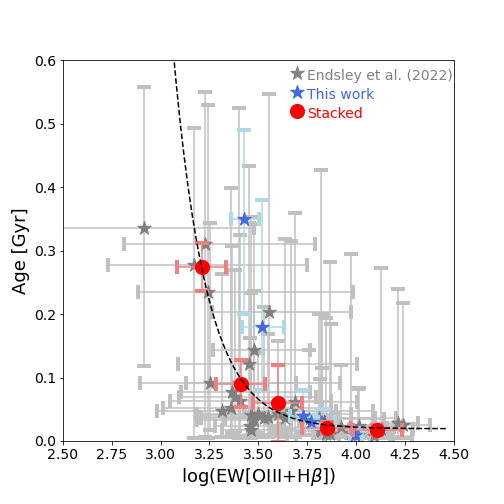}
    \caption{Distribution of the equivalent width of [OIII]+H$\beta$ (estimated from the NIRCam photometry) as a function of the age of the stellar population (estimated by \texttt{BAGPIPES}) for galaxies in the UDF medium-band survey (blue). We compare this distribution with that at lower redshift $z\sim$6.8 by \citet{2021MNRAS.500.5229E}. We also averaged the measures in bins of EW[OIII+H$\beta$] (red). The resulting trend confirms that extreme [OIII]+H$\beta$ emission is a signature of young stellar populations, but galaxies with EW[OIII+H$\beta$]$<$10$^3$ can nonetheless harbour an older stellar population. Error bars for the "stacked" data-points are the standard deviation of the values.}
    \label{fig:age-vs-OIII}
\end{figure}

Medium-band photometry also improves the accuracy of the inferred equivalent width (EW) of [OIII]5007,4959\AA +H$\beta$. It has been a long-standing debate as to whether or not the strength of [OIII]5007,4959\AA can indicate the age of the stellar population. Our estimate of the strength of the Balmer break from the stellar continuum traced with the medium-band filters allows us to re-examine this question. Figure~\ref{fig:age-vs-OIII} shows the age of the stellar population as a function of the EW(OIII+H$\beta$) for our sample (blue dots) as compared with a similar analysis conducted at lower redshift ($z\sim$6.8 - \citealt{2021MNRAS.500.5229E}). We average the data in bins of EW[OIII+H$\beta$] and take the standard deviation as indication of the uncertainty (red dots). The figure confirms the natural suspicion that extremely strong EW[OIII+H$\beta$] emission is only seen for young stellar population but, importantly, that less intense lines are still consistent with the presence of older stellar populations. We fit the binned data with an exponential curve and obtain the following empirical relation :
\begin{equation}
    Age[Myr]= A \times \exp{(-B * \log (EW_{OIII+H\beta})}) + C
\end{equation}
with $A=3.64\times10^{10}$ , $B=5.85$ and $C=19.4$. 

The final physical property we can estimate from our SED-fitting analysis is the stellar metallicity. The mean metallicity of our sample is $<Z>$=0.62$^{+0.26}_{-0.21}$ Z$_{\odot}$. 
Our values are consistent with estimates derived using ALMA [OIII]88$\mu$m luminosities \citep{2020ApJ...903..150J}. They are also in good agreement with that estimated for a spectroscopically confirmed $z=9.76$ galaxy, $Z_{\star}$=0.59$\pm$0.12 $Z_{\odot}$ \citep{2022arXiv221015639R}.

\begin{table*}
    \centering
    \begin{tabular}{ | l | cccccc | }
    \hline
ID 	&	z	&	$\log$[M$^{\star}$]	&	SFR	&	Age	& EW[OIII+H$\beta$]	& Z \\ 
&   &   M$_{\odot}$ & M$_{\odot}$/yr & Gyr  & \AA\ & Z$_{\odot}$\\ \hline
UDF-s2-1	&	8.21$^{a}$ $^{+ 0.03 }_{- 0.04 }$	&	9.19 $^{+ 0.07 }_{- 0.06 }$	&	15.60 $^{+ 2.72 }_{- 1.87 }$ &	0.03 $^{+ 0.01 }_{- 0.01 }$	&	5900$^{+514}_{-473}$ &	0.94 $^{+ 0.13 }_{- 0.18 }$ \\
UDF-s2-2	& 8.27 $^{+ 0.02 }_{- 0.08 }$	&9.26 $^{+ 0.09 }_{- 0.09 }$ &	9.06 $^{+ 2.31 }_{- 1.50 }$	&	0.35 $^{+ 0.14 }_{- 0.15 }$	&	2680$^{+461}_{-416}$ &	0.52 $^{+ 0.26 }_{- 0.23 }$ \\
UDF-s2-3	&	8.19$^{b}$ $^{+ 0.04 }_{- 0.05 }$	&	8.90 $^{+ 0.13 }_{- 0.16 }$	&	8.12 $^{+ 3.09 }_{- 2.24 }$	&	0.04 $^{+ 0.04 }_{- 0.02 }$	&	5339$^{+859}_{-749}$ &	0.72 $^{+ 0.36 }_{- 0.27 }$ \\
UDF-s2-4	&8.24 $^{+ 0.02 }_{- 0.04 }$ &	8.78 $^{+ 0.13 }_{- 0.12 }$	&	6.38 $^{+ 1.75 }_{- 1.42 }$	&	0.03 $^{+ 0.02 }_{- 0.01 }$	&	6894$^{+887}_{-782}$ &	0.46 $^{+ 0.27 }_{- 0.17 }$ \\
UDF-s2-5	&	8.14 $^{+ 0.06 }_{- 0.06 }$	& 8.97 $^{+ 0.14 }_{- 0.17 }$	&	6.49 $^{+ 1.50 }_{- 1.29 }$	&	0.18 $^{+ 0.20 }_{- 0.10 }$	&	3316$^{+883}_{-740}$ &	0.52 $^{+ 0.28 }_{- 0.21 }$ \\
UDF-s3-1	&	8.69 $^{+ 0.03 }_{- 0.03 }$	&	8.75 $^{+ 0.08 }_{- 0.07 }$	&	5.70 $^{+ 1.20 }_{- 0.89 }$	&	0.01 $^{+ 0.01 }_{- 0.00 }$	&	9848$^{+553}_{-769}$ & 0.54 $^{+ 0.23 }_{- 0.15 }$	\\

 \hline
    \end{tabular}
    \caption{Physical properties of our [OIII]5007\AA\ emitters. The photometric redshift, stellar mass, SFR, age and metallicity have been computed with \texttt{BAGPIPES}, whereas the equivalent widths of [OIII]+H$\beta$ are estimated from the NIRCam photometry. The good precision for the photometric redshifts ($\Delta z_{phot}$=0.08) and stellar masses ($<\Delta \log [M_{\star}]>$=0.21) demonstrate the advantage of using medium-band filters. \\
    $^a$ : spectroscopically confirmed at $z=$8.228 \\
    $^b$ : spectroscopically confirmed at $z=$8.224}
    \label{tab:properties}
\end{table*}

\section{Discussion}
Prior to the first observations with the \textit{James Webb} Space Telescope, the optimal technique used to estimate the age of distant galaxies utilised the combination of \textit{Hubble} Space Telescope photometry with \textit{Spitzer} fluxes at 3.6 and 4.5$\mu$m. However, because of contamination of the \textit{Spitzer} photometry by [OIII]4959,5007\AA and H$\beta$ emission lines, ambiguities of interpretation could only be avoided for galaxies at $z\geq$9.1 where [OIII]5007 is redshifted beyond the 4.5$\mu$m band (eg, \citealt{2018Natur.557..392H}, \citealt{2021MNRAS.505.3336L}). Adding ALMA constraints on the [OIII]88$\mu$m line flux may be used to constrain the strength of [OIII]5007\AA \citep{2020MNRAS.497.3440R}, but very few $z\geq$8 sources currently have both [OIII]88$\mu$m detections and Spitzer photometry. The question remained, therefore, whether or not galaxies with evolved stellar populations are common at $z\sim$7-9. Using \textit{Hubble} and \textit{Spitzer} data, \citet{2022ApJ...927..170T} selected 11 galaxies at $z\geq$8, and concluded that only one potentially hosted an evolved stellar population (UDS-18697). For our $z\geq$8 medium-band sample, UDF-s2-2 and UDF-s2-5 indicate a mature ($\geq$150 Myr) stellar population with an age of 350$^{+140}_{-150}$ and 180$^{+200}_{-100}$ Myr respectively corresponding to a formation redshifts of
$z_{form}$=15.15$^{+10.25}_{-4.12}$ and 10.43$^{+5.42}_{-1.43}$. Although a modest sample, 1/3 of our strong [O III] emitters may thus have formed at $z\geq$10. 



Finally, the improved sensitivity of JWST imaging allows us to study the environment of luminous objects at high redshift. Two galaxy protoclusters have already been identified at $z\geq$7.5 (\citealt{2022A&A...667L...3L}, \citealt{2022arXiv221109097M}) with several spectroscopically-confirmed members. Three galaxies from our $z\sim$8.2 sample (UDF-s2-1, UDF-s2-2 and UDF-s2-3) are co-located on the same NIRCam detector with a projected separation of 31\arcsec between UDF-s2-1 and UDF-s2-2 (corresponding to 0.15 Mpc), 1.18\arcmin (0.34 Mpc) between UDF-s2-1 and UDF-s2-3 and 1.01\arcmin (0.29 Mpc) between UDF-s2-2 and UDF-s2-3. However, these separations are larger than those observed in SMACS0723 protocluster (40 $\arcsec$ - 0.2Mpc) and behind A2744 ($<$50 kpc). Moreover the luminosities are fainter than in previous findings ($m_{F182M} \geq$27.5 AB) suggesting a physical association is unlikely. 

\section{Summary and Conclusions}

In this letter, we searched for strong [OIII]5007 \AA\ emitters at $z\geq$8 in the first medium-band survey undertaken with the \textit{James Webb} Space Telescope. The detection of a strong emission line in one of the medium-band filters leads to an accurate photometric redshift ($\Delta z_{phot}$=0.08). Over the 9.68 arcmin$^2$ NIRCam field, we located 6 $z\geq$8.2 candidates with F182M magnitudes ranging from 27.8 to 28.7 AB. For several of these we confirm the redshifts spectroscopically using NIRCam grism data from the FRESCO survey. Although the remaining sources might be contaminated by strong H-$\alpha$ emitters at $z\geq$6, we argue this is unlikely given the absence of a detection in a deep ACS stack and because this would imply a large dust content ($A_v \geq$2mag) inconsistent with recent results from a deep ALMA band 6 survey.  Using a range of assumed star formation histories, we demonstrate that a third of our sample likely host an established stellar population (age$\geq$150 Myr) whose formation redshifts would be $z_{form} \geq$10.  
Our main conclusion that a significant fraction of intense [OIII] emitters contain underlying mature stellar populations is clearly consistent with the recent detection with JWST of $z\geq$15 galaxies (eg. \citealt{2022ApJ...938L..15C}, \citealt{2022arXiv220712356D}). The precise location of the [OIII]5007\AA\ emission line in the relevant medium-band filter and the accurate measurement of the stellar continuum also leads to improved estimates of the age and stellar mass ($<\Delta \log [M_{\star}]>$=0.21). By comparing our results with those at lower redshift, we demonstrate a logical trend between the strength of the [OIII]+H$\beta$ equivalent width and the age of the stellar population.  

Our study demonstrates the important role of medium-band JWST filters in determining accurate physical properties of early galaxies, as expected from simulations \citep{2021ApJ...910...86R}. Although a modest sample based on data over a small field of view ($<$10 arcmin$^2$), more extensive imaging would enable much progress in unraveling the past star formation histories of $z\sim$7-9 galaxies.  

\section*{Acknowledgements}

We thank Angelica Lola Danhaive and Fengwu Sun for their help in the NIRCam/grism data reduction. Authors thank Christina Williams, Michael Maseda, and Sandro Tacchella, the co-PIs of the JWST program (PID 1963) we used in this study.  NL acknowledges support from the Kavli fundation. RSE acknowledges funding from the European Research Council (ERC) under the European Union’s Horizon 2020 research and innovation programme (grant agreement No 669253). CECW thanks the Science and Technology Facilities Council (STFC) for a PhD studentship, funded by UKRI grant 2602262. This work is based on observations taken by the 3D-HST Treasury Program (GO 12177 and 12328) with the NASA/ESA HST, which is operated by the Association of Universities for Research in Astronomy, Inc., under NASA contract NAS5-26555.

\section*{Data Availability}

Reduced images and catalogues will be shared on reasonable request to the corresponding author.



\bibliographystyle{mnras}
\bibliography{mnras_template} 




\appendix
\begin{landscape}
\bsp	
\label{lastpage}
    \setcounter{table}{1}
\begin{table}
\begin{center}
\scriptsize
    \begin{tabular}{ | l | ll | cccccccccccccc | }
    \hline
ID	&	RA	&	DEC	&	F435W	&	F606W	&	F775W	&	F814W	&	F850LP	&	F105W	&	F125W	&	F140W	&	F160W	&	F182M	&	F210M	&	F430M	&	F460M	&	F480M	\\ \hline
UDF-s2-1	&	03:32:35.02	&	-27:49:22.1	&	$>$28.7	&	$>$29.0	&	$>$28.2	&	$>$28.0	&	$>$28.3	&	$>$28.5	&	$>$28.0	&	$>$27.2	&	27.82$\pm$0.18	&	28.58$\pm$0.06	&	27.58$\pm$0.03	&	26.82$\pm$0.05	&	25.29$\pm$0.02	&	26.53$\pm$0.04	\\
UDF-s2-2	&	03:32:32.70	&	-27:49:20.6	&	$>$28.7	&	$>$29.0	&	$>$28.2	&	$>$28.0	&	$>$28.3	&	28.48$\pm$0.17	&	28.41$\pm$0.19	&	$>$27.2	&	28.29$\pm$0.28	&	27.83$\pm$0.03	&	27.78$\pm$0.04	&	27.06$\pm$0.06	&	26.1$\pm$0.04	&	26.82$\pm$0.05	\\
UDF-s2-3	&	03:32:32.02	&	-27:50:20.7	&	$>$28.7	&	$>$29.0	&	$>$28.2	&	$>$28.0	&	$>$28.3	&	$>$28.5	&	$>$28.0	&	$>$27.2	&	$>$27.5	&	28.25$\pm$0.05	&	27.96$\pm$0.05	&	27.22$\pm$0.08	&	25.86$\pm$0.04	&	27.1$\pm$0.07	\\
UDF-s2-4	&	03:32:30.74	&	-27:49:45.0	&	$>$28.7	&	$>$29.0	&	$>$28.2	&	$>$28.0	&	$>$28.3	&	$>$28.5	&	$>$28.0	&	$>$27.2	&	$>$27.5	&	28.05$\pm$0.04	&	27.95$\pm$0.05	&	27.24$\pm$0.08	&	25.58$\pm$0.03	&	26.94$\pm$0.06	\\
UDF-s2-5	&	03:32:30.63	&	-27:49:03.3	&	$>$28.7	&	$>$29.0	&	$>$28.2	&	$>$28.0	&	$>$28.3	&	29.04$\pm$0.26	&	28.82$\pm$0.26	&	$>$27.2	&	$>$27.5	&	28.62$\pm$0.09	&	28.26$\pm$0.09	&	27.5$\pm$0.09	&	26.58$\pm$0.06	&	27.59$\pm$0.10	\\
UDF-s3-1	&	03:32:39.87	&	-27:46:19.5	&	$>$28.7	&	$>$29.0	&	$>$28.2	&	$>$28.0	&	$>$28.3	&	$>$28.5	&	$>$28.0	&	$>$27.2	&	$>$27.5	&	27.96$\pm$0.04	&	27.77$\pm$0.04	&	26.92$\pm$0.06	&	26.55$\pm$0.06	&	25.25$\pm$0.01	\\

    \end{tabular}
    \end{center}
    \caption{Photometry of the selected candidates. UDF-s1-* refers to the $z\sim$7.6 sample, UDF-s2-* refers to the $z\sim$8.2 sample and UDF-s3-* to the $z\sim$8.6 sample. Magnitudes are the total MAG\_AUTO magnitudes, upper limits are at 2$\sigma$. }
    \label{tab:photometry}
\end{table}
\end{landscape}



\end{document}